\documentclass{article}
\pdfoutput=1
\usepackage{amsfonts}
\usepackage{graphicx}
\usepackage{authblk}
\newcommand{\beginsupplement}{%
        \setcounter{table}{0}
        \renewcommand{\thetable}{S\arabic{table}}%
        \setcounter{figure}{0}
        \renewcommand{\thefigure}{S\arabic{figure}}%
     }

\title{Mixture model of pottery distributions from Lake Chad Basin archaeological sites reveals ancient segregation patterns}

\author[1]{John D. O'Brien\thanks{jobrien@bowdoin.edu}}
\author[1]{Kathryn Lin}
\author[2]{Scott MacEachern}
\affil[1]{Department of Mathematics, Bowdoin College}
\affil[2]{Department of Sociology and Anthropology, Bowdoin College}

\begin{document}

\maketitle

\section*{Abstract}
We present a new statistical approach to analyzing an extremely common archaeological data type -- potsherds -- that infers the structure of cultural relationships across a set of excavations. This method, applied to data from a set of complex, culturally heterogeneous sites around the Mandara mountains in the Lake Chad Basin, articulates currently understood cultural succession into the Iron Age. We show how the approach can be integrated with radiocarbon dates to provide detailed portraits of cultural dynamics and deposition patterns within single excavations that, in this context, indicate historical ethnolinguistic segregation patterns. We conclude with a discussion of the many possible model extensions using other archaeological data types.



\section*{Introduction}
A central goal of archaeology is to infer the variety of cultures that were present at a given location and their relative succession in time. These estimates preferably come with corresponding ages and take account of the fact that such cultures are complex analytical entities. Traditionally, technical innovation in archaeology toward this aim has largely originated in the physical sciences, such as radiocarbon dating (RCD), chemical composition analysis, and complex microscopy \cite{Browman1996,Tite1974,Mantler2000}. Recently the discipline begun to integrate the power of computational statistics, though this use has largely been limited to contemporary genetic and linguistic data \cite{Hanotte2002,Gray2009,Tishkoff2009,Nicholls2001}. In this work, we fuse these trends by applying a well-known computational statistical methodology to one of the most common archaeological data -- potsherds -- to infer cultural temporal dynamics.

\begin{figure}[!ht]
\begin{center}
\includegraphics[scale=0.5]{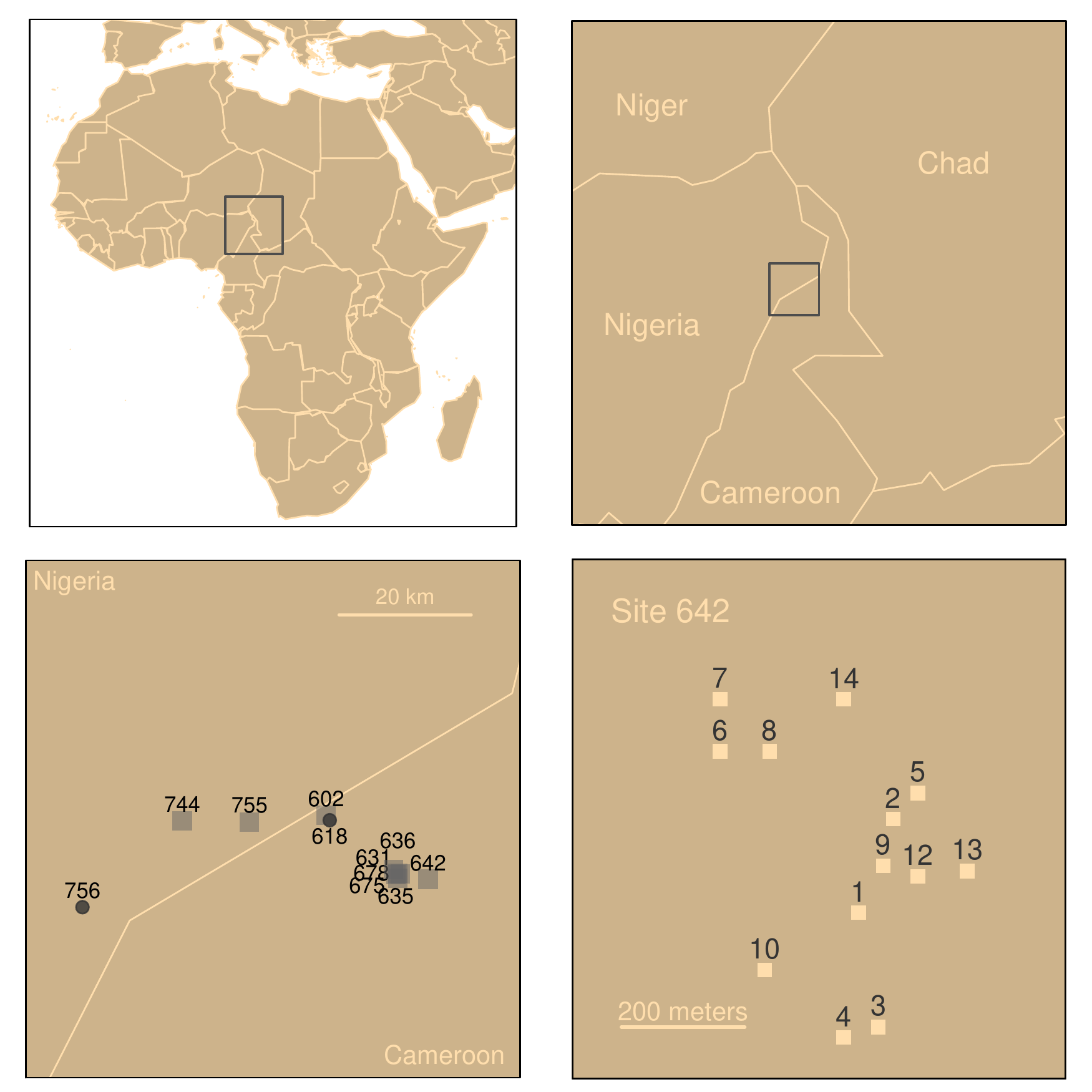}
\end{center}
\caption{Location of the eleven field sites on the Cameroon-Nigeria border to the south of Lake Chad near the Mandara Mountains. The two sites marked as dark grey dots -- 618 and 756 -- possess Neolithic features. Lower right panel shows the relative position of excavation units (EUs) within site 642. Maps constructed using the \emph{ggmap} library \cite{Kahle2013}.}
\label{locations}
\end{figure}

Here we consider a data set comprised solely of potsherds derived from eleven Neolithic and Iron Age sites in the southern Lake Chad Basin of Central Africa, and classified using standard anglophone descriptive nomenclature \cite{Haour2010}. Ethnolinguistic diversity among the populations in this area (Figure \ref{locations}) is among the highest in Africa  and comparable to anthropologically well-known regions like highland New Guinea. This diversity is particularly great in and around the Mandara Mountains on the Cameroon-Nigeria border, occupied at high population densities by Chadic-speaking \emph{montagnard} farmers. The historical roots of that diversity are complex:  after the retreat of Lake Mega-Chad populations speaking ancestral Chadic and Nilo-Saharan languages reoccupied the region in the mid-Holocene \cite{Schuster2005,Bouchette2010,Tishkoff2009} with associated sedentism, population growth and cultural diversification on the plains to the south	 between 4000 and 1200 BP. Progressive occupation of the nearby Mandara Mountains by plains groups occurred as political conflict and slave raiding became regionally significant \cite{MacEachern2012}, with the deployment of linguistic diversity and multilingualism as sociolinguistic strategies in a challenging mountain environment \cite{Sankoff1969}.

Between 1984 and 2003, the Mandara Archaeological Project investigated the material correlates of this modern ethnolinguistic diversity  \cite{David2012,Lyons1996,David2012b}. As throughout much of Africa, ceramics in the Mandara Mountains are a vital material domain for the expression of identities: domestic pottery, elaborated both in morphology and in decoration, is ubiquitous and central to social interactions across societies \cite{David1988,vanBeek2012}. In the modern Mandara context, there are strong connections between ceramic variability and the ways its production and use are associated with identity. Simple concordance may be confounded by the sharing of ceramic traditions among groups \cite{MacEachern1998},  possibly due to women frequently marrying across ethnolinguistic boundaries and carrying their habits of pottery production with them \cite{Richard1977,Lyons1992,Gosselain2002}. Nonetheless, we expect ceramic variability still largely reflects cultural traditions and long-standing social interactions both at single sites and across larger geographic scales.

Archaeological research indicates that modern Mandara populations appear to be related historically and culturally to Iron Age and Neolithic communities living on the plains adjacent to the mountains \cite{MacEachern2012}. These sites date to between approximately 3500 BP and the historical period. It is likely that sites from at least the last 2000 years are associated with Chadic-speaking farming populations. As in many equivalent Neolithic and Iron Age West African sites, pottery was recovered in very large quantities, with 312,000 potsherds comprising well over 99\% of artifact inventories. As a result of the conditions of preservation of this material, very few whole pots were excavated and most potsherds are quite small. Consequently, morphological data are often not available and the primary attribute available for analysis is external decoration. As the modern context suggests a strong coupling of ceramic variability and identity, we assume that this is also true within the archaeological data. The statistical approach here then seeks to identify patterns of similar decoration distribution to infer ancient group identities.

A longstanding archaeological dilemma is how to address the incomplete mapping of physical artifacts to their associated cultures as important shifts may not be detected by a single data type\cite{MacEachern2012b}. We consequently employ the term ‘cultural period’ (CP) to denote a consistent distribution of potsherd decoration types. This means that CP distributions may shift due to migration, marriage patterns, trade, other social interactions, or technological diffusion, rather than the signifying the emergence of new cultures. 		

The core of the analysis is a Dirichlet process mixture model (DPM), a standard Bayesian nonparametric approach to mixture modeling, applied to distributions of decoration classifications from potsherd collections \cite{Antoniak1974,Ferguson1983,Gelfand2002}. These models are widely used in statistics, computer science, and biology in contexts where both the underlying number of model components and their parameters are unknown \cite{Lartillot2004,Muller2004,Teh2006,Xing2006,Huelsenbeck2007}. Here, CPs define the underlying components, with each CP corresponding to a distribution of potsherd decorations during that period. The statistical framework then provides a means to infer the CPs and their underlying parameters from the data, giving a generative model of the cultural dynamics underlying potsherd production across sites and depths.

The process of achaeological excavation necessitates a choice about how to best group the data for analysis by the model. Most of the field sites are subdivided into excavation units (EU) that are in turn broken down into meter-square-sized recording units (RU) for material recovery. At each 10 cm depth increment within a square removed archaeological material is collected for sorting, cleaning, classification, and further analysis; consequently, this is the most particularized form that the decoration data can take. On the one hand, data aggregated at the RU-level often contained too few potsherds to provide much inferrential power, while aggregations at the site level often masks interesting variation. The analysis presented here is based on the EU-level aggregations, as this preserved the necessary variation while providing sufficient counts for the algorithm to work reliably. The data for each depth we call unit-levels. The output of the algorithm we refer to as a culture painting.

\section*{Data and Methods}

\subsection*{Data and notation} \ The complete data set is made up of $312,070$ archaeological potsherd observations, collected over 31 years of field collection at 80 dig sites in the Mandara region to the south of Lake Chad on the Cameroon-Nigerian border \cite{David2012}. Following standard classification procedures, these potsherds were categorized according to external decoration type, potsherd type, and interior decoration type, along with estimates of rim diameter and vessel type, where possible \cite{Jones2001}. The statistical method presented here uses only external decoration data for three reasons: these were available for the vast majority of records; they exhibited significant variation across and within sites; and they provided significant information to previous researchers. 

The data set contains 86 external decoration types, each possessing between one and 72,100 records, with 361 associated subtypes. We filtered the data set to contain only decoration types with at least 2,000 observations to ensure sufficient frequency for statistical inference. We also removed sites with fewer than one hundred observations, leading to eleven final sites. The final data set includes 239,629 potsherds (76.8\% of the total). The final set of sites are listed according to the site designation system used for the wider archaeological project: sites 602, 618, 635, 636, 642, and 675 for those in Cameroon and 744, 755 and 756 for those in Nigeria. Most sites (7/11) contain multiple EUs, that in turn often include multiple RUs. For instance, site 642 contains 14 EUs, each of which included multiple RUs. Within each EU, stratigraphic levels correspond to depths of sediment removed for analysis, measured in 10 cm intervals beginning from the ground surface, with the total number of levels ranging from 0 to 33. Correspondingly, each potsherd is then indexed by its unit $i = 1, \cdots, 36$, level $j=1\cdots,34$, and decoration type $d=1,\cdots,13$. We use $s_{ij}$ to refer to $(s_{ij,1}, \cdots,s_{ij,13})$, the collection of external decoration counts for unit-level $(i,j)$. 
\subsection*{Model} \ The model assumes that the unit-level counts arise from an unobserved CP that governs the production of potsherd decorations. We assume CP $k$ to have the form of a Dirichlet-multinomial (DM) distribution parameterized by $A_k = (\alpha_{k,1}, \cdots, \alpha_{k,13})$, where $\alpha_{i,k}$ is the concentration parameter for the $i^{\mbox{\tiny{th}}}$ decoration type. A latent variable $c_{ij}$ associates each unit-level (i,j) to a CP. Conditional upon $c_{ij}=k$, the likelihood at the unit-level is then:
\begin{eqnarray}
\mathbb{P}(s_{ij}| c_{ij}, A_k ) &=& \frac{\Gamma(\overline{A}_k)}{\Gamma(\overline{N}_{ij}+\overline{A}_k)} \prod_{d=1}^D \frac{\Gamma(s_{ij,d} + \alpha_{k,d})}{\Gamma(\alpha_{k,d})}  \label{likelihood} \\
& = & \mbox{\small{DM}}(A_k)   \nonumber
\end{eqnarray}
where $\overline{A}_{k} = \sum_{d=1}^D \alpha_{k,d}$, and $\overline{N}_{ij} = \sum_{d=1}^D s_{ij,d}$. This likelihood allows us to account for greater dispersion in the data than would be possible with a strict multinomial distribution. The unit-levels are conditionally independent given their CP assignment and the CP parameters so it remains only to specify their prior distributions to complete the model.

Since the total number of CPs is not known \emph{a priori}, we turn to the DPM, a Bayesian nonparametric approach to prior specification, for a distribution on the number of CPs present and the corresponding DM parameters. Following \cite{Neal2000}, the DPM with DM components can be formulated as:
\begin{eqnarray*}
s_{ij}|c_{ij} = k	 & \sim & \mbox{\small{DM}}(A_k) \\
A_k | G & \sim & G \\
G &\sim & \mbox{\small{DP}}(G_0,\gamma) 
\end{eqnarray*}
where $G_0$ is the base measure, $\gamma>0$ is a concentration parameterm, $\mbox{\small{DM}}(A_{c_{ij}})$ denotes the likelihood in Equation \ref{likelihood}, and $\mbox{\small{DP}}$ is a Dirichlet process. The base measure $G_0$ is the product of two independent distributions, with the first component an exponential distribution with mean one and the second a uniform Dirichlet distribution, that is chosen for partial conjugacy with the DM likelihood \cite{Stein2013}.

\subsection*{Inference} \ We employ standard Markov chain Monte Carlo (MCMC) approaches to sample from the posterior distribution of the DPM to infer the number of CPs, the parameters $A_k$ associated with each CP, and the assignment of unit-levels to specific CPs \cite{Neal2000}. The implementation uses established algorithms for inference under DPMs, together with an efficient Gibbs sampling routine specific to the DM distribution \cite{Stein2013}. Since the assignment of unit-levels to CPs is not unique up to labeling, we employ a modified version of a Kullback-Leibler minimization method to re-label the posterior samples to align CPs for each iteration with the maximum likelihood iteration \cite{Stephens2000}. Ten independent runs of the algorithm with one million iterations provide nearly identical results, both in terms of number of cultures and inferred DM distributions. The run presented in the Results section has an effective sample size of 265.31, indicating sufficient sampling for reliable inference. A complete outline of the computational approach is outlined in the Supplementary Information. Along with the cleaned data set, an implementation of the model in the R computing environment is available at the Digital Archaeological Record \cite{Team2000}.
\section*{Results}
Figure \ref{painting} shows the culture painting for the Mandara data set. This presentation uses 14 primary CPs identified via an incidence plot (Figures S1 and S2), that we enumerate C1-C15, with CP 15 representing an aggregation of all clusters outside of the primary 14. Coloring indicates the fraction of posterior samples assigned to a CP over the posterior sample. Figure S3 shows inferred potsherd distribution for each culture and a hierarchical clustering of their relatedness, with darker shading equating to more data. Figure S4 shows Figure \ref{painting} without shading. 

The culture painting shows signficant variability in CP assignment across sites, EUs, and depth, as would be anticipated by previous regional archaeological and linguistic analysis. This heterogeneity in CPs across sites contrasts with strong consistency in CP assignment within the same site (for example, sites 675 and 756) that conveys settlement continuity, while long runs of identical CPs across successive depths with occassional shifts (for instance, site 602 EU 2 and site 642 EU 2) indicate discernible shifts in production patterns. The observation of nearly all CPs across multiple sites (only CPs 2 and 3 are seen solely within a single site) and successive depth indicate the model possesses sufficient power to infer similar cultural patterns across space and time.

\subsection*{Culture painting preserves geographic separation } \ The physical separation of the western sites (602, 618, 744, 755, and 756) compared to the more easterly sites (631, 635, 636, 675, and 678) is largely preserved in the CP assignments, with the western locations dominated by CPs 1, 5, 8, 9, 10, and 13 with smaller contributions from CPs 2, 3 and 14, while the main sites largely exhibit CPs 1, 3, 4, 6, 7, 8, 10, 11, 12 and 14 with smaller amounts of CP 9. The overlapping CPs (8 and 10) between these two collections of sites largely appear at greater depths within the easterly sites. This segregation supports a long-standing geographic/cultural discontinuity between western and eastern parts of the research areas, over the last 2000 years at least.

\subsection*{Distinction of Neolithic sites} \ Through RCDs and ceramic affiliations, sites 618 and 756 were established to be of the Neolithic period, and are associated predominantly with CP 14, with site 756 nearly exclusively so \cite{Ramsey2001}. The potsherd distribribution for CP 14 is one of the more distinct of those inferred, as would be expected for earlier and very different cultural and material adaptation (Figure S3). The algorithm also associates additional unit-levels at other sites with CP 14 that, consistent with the hypothesis of these coming from an earlier culture, occur largely at the greatest depths within EUs, as in sites 631 EU 1, 636 EU 4, 642 EU 4. These deeper unit-levels may be associated with a Neolithic-to-Iron Age cultural transition occurring about 2500 years ago. Exceptions occur within two sites, 744 and 755 (both located near site 756), with unit-levels associated with CP 14 at near-surface unit-levels, indicating that either these EUs may be partially Neolithic or that taphonomic processes introduced deposits containing Neolithic sherds into later stratigraphic contexts.

\subsection*{Uneven transitions} \ Despite the high correlation among successive depths, many sites also exhibit repeated switching between two CPs, as exemplified by site 602 EU 2, site 635 EU 3, and site 642 EU 2. These `switchbacks' may indicate slow, inconsistently progressive cultural shifts, potsherd movement between levels, or aggregation of culturally-distinct excavation units. Simulation results described below indicate that mixture due to potsherd migration would need to be extensive to entirely explain these effects. Consideration of the culture painting for the RU unit-level aggregation suggests that some fraction of these events may be attributed to aggregation effects (see, for instance, site 642 in Figure S8). The remaining events appear to represent shifts within the data, suggesting that the algorithm is in certain cases sufficient to resolve cultural dynamics at previously inaccessible levels  of granularity. These will need to be further studied by broader comparisons with other, less common data types, including stone tools and clay figurines, but those comparisons are beyond the boundaries of the present analysis.

\subsection*{RCDs at site 642 indicate ancient segregation} \ The culture painting for site 642, the most complex and extensive excavation in the region, also exhibits the most complex patterning, with evidence of significant internal differention among EUs. This site is comparable in size to modern farming villages in the region, and such villages often include ethnically-segregated neighborhoods. The algorithm indicates that the units can be organized into five broad groups: (1) EUs 1, 1B, 1C that exhibit CPs 4 and 9 near the surface, transitioning to CPs 3 and 8 at greater depths; (2) EU 2 has repeated successive shifts between CP1 and CP2 with progression to CP 10 at 2.2 meters ; (3) EUs 3 and 4 associate with CP 11 from the surface down to approximately 1.4 meters, then change largely into CP 8; EU 5 shows CP 8 changing to CP 9 and then CP 1; and EUs 6-14 show CPs 3 and 11, with introgressions from CP 8 and significant posterior uncertainty consistent with their low number of counts. Site 631 shows similar but less eleborate segregation patterns between EUs 1 and 2 and EU 3. These stable but distinctive patterns within each group support the previously hypothesized propsition that modern ethnic segregation within Lake Chad Basin villages is inherited from deeper historical divisions \cite{MacEachern2012}. 

Ancient segregation patterns are also supported by integrating RCD information with the culture painting for the subset of EUs with that information, as shown in Figure \ref{rcd}. EUs 3 and 6 exhibit distinct CPs from both EUs 1 and 2 and from each other for depths dated to approximately 1000 years before the present, with the culture painting showing consistent CPs for neighboring unit-levels. EUs 1 and 2 also have largely distinct CPs patterns from the surface down to the dated depths, consistent with the continuity of segregation patterns into the historical period. However, CP2's substantial but variable presence in both EUs suggests more complex cultural dynamics between these two locations than strict segregation.

\subsection*{Simulations}  \ To ensure that model inference is reliable, an additional simulation study was performed to understand how the total number of potsherds, the number of decoration types, the number of unit-levels, the degree of autocorrelation among levels, and the amount of potsherd mixing across levels contribute to algorithm performance. Data was simulated as follows. The number of CPs was fixed to five and the number of unit-levels was fixed to one hundred and organized into five sites with twenty unit-levels. The remaining parameters were varied for each simulation. Unit-levels were assigned randomly to a CP with a probability $\rho$ of switching cultures between successive unit-levels. For each CP, a set of parameters were randomly generated to determine the external decoration distribution, with the concentration parameters sampled independently from an exponential distribution with mean one. Conditional upon the total number of counts, decoration data were then generated for each unit-level. To simulate potsherd mixing, a fraction $f$ of each unit was swapped with neighboring unit-levels.  A complete description of the parameters used are given in Table S1. Each set of parameters was repeated for ten iterations. 

As shown in Figure S9, we find that the algorithm's performance depends largely on the total number of potsherds, the number of decoration types, and, to a lesser degree, the amount of mixing between unit-levels, as measured by three metrics: the Kullback-Liebler divergence between the maximum-likelihood posterior sample and the simulated CP assignment; the mean correlation between $A_c$ for inferred and simulated CPs; and the inferred number of CPs. Performance is broadly consistent for each metric, with little improvement in performance when there are more than 15 decoration types and 250 potsherds within a unit-level. Substantially below these levels inference can be highly variable. The amount of mixing only noticeably affects inference at very high rates (50\% of potsherds) and with low amounts of unit-level correlations. As would be expected from the absence of correlation modeling, performance is insensitive to the degree of autocorrelation among sites when mixture is not considered. Within the Mandara data set we have 13 decoration types, with a median number of counts of 325 (7 and 1567 for 5\% and 95\% values, respectively), placing the large majority of unit-levels safely within the reliable inference regime.

\section*{Discussion}

The model presented provides a new approach to modeling distributions of ancient artifacts, admitting quantitative comparison about cultural affiliations, their successional dynamics, and the distribution of their material production. Inferred from potsherd data collected at a set of sites ranging across a culturally complex region, the model provides insight that is both consistent with current research and provides significant extensions to current practice, demarcating cultural shifts as well as the uncertainty in these estimates. This analysis indicates that aspects of cultural patterning seen today (especially in terms of the spatial distribution of ethnolinguistic groups, and cultural segregation within nucleated villages) may have deep-time antecedents extending more than 1000 years back into the Iron Age. This is consistent with overall relationships between modern and Iron Age ceramic assemblages in the region. Further, this provides support for a qualitative assessment of differences between Neolithic and Iron Age ceramic assemblages, and for the existence of transitional assemblages in the mid-third-millennium BP. This approach will likely be useful in exploring other cultural processes in the region, including the Iron Age occupation of the Mandara Mountains themselves. Decoration data can also be compared to other regions of Africa in order to examine cultural dynamics over much larger spatial scales.
 
This modeling approach is not limited to potsherd decoration and may be usefully extended to a variety of artifactual data. In particular, the model only leverages one form of classification, while the majority of potsherds have additional categorization based on vessel type, potsherd type, and potsherd location within the vessel. These data provide complementary information to external decoration so model extensions that include them may provide increased resolution of cultural shifts. At the other extreme, the insistence on high-frequency decoration types in this model also excludes exotic potsherds and decoration types that researchers have traditionally used to infer long-range trade patterns, such as the sgraffito decoration type often held to indicate Kanuri presence or contact \cite{Peacock1969}. The Bayesian formulation of the model holds out the possibility that these data and other non-ceramic forms of data -- notably linguistic, genetic, mineral trace information, and other artifact types (lithics, metals) -- could be integrated to provide rich frameworks for data analysis and hypothesis testing.  

The integration with RCDs presents a point of especial interest to archaeology, as it connects physical depths into with absolute dates. As shown in Figure \ref{painting}, the model's inference provides a sequential understanding of cultural patterning within each EU that is not always consistent within sites. Extending the model's statistical scope to include these data and map all unit-levels onto a uniform time scale would further enhance the portrait of ancient taphonic and cultural dynamics. This would also provide a new method for understanding the interaction of settlements, soil types, and deposition rates, potentially between and within sites. However, this extension presents signficant statistical challenges since bringing conflicting RCDs and unit-levels that lack RCDs into correspondence has no ready solution.

The analysis undertaken here deliberately does not model correlations among neighboring unit-levels or across EUs within the same site. This is partially to understand the sensitivity of the model to detect distributional shifts, but also to avoid the perils of overfitting. However, modeling these correlations provides the promise of `borrowing strength' across EUs to impute CPs where data is scarce, as is commonly possible in Bayesian analysis \cite{Best2005}. The specification of infinite hidden Markov models, particularly allowing for variable dwelling times, may prove a fruitful extension \cite{Dewar2012}. The generation of parameters for the DM distribution may also be extended by modeling how more recent cultures arise from older ones according to, for instance, a hierarchical Dirichlet process, or by providing additional parameteric flexibility afforded by a logistic multivariate normal distribution \cite{Minno2008}.

An operational question raised by this study is how researchers should aggregate potsherd data for analysis by the culture painting algorithm. As noted in the introduction, in the Mandara mountain excavation most sites contain multiple EUs that in turn often contain multiple RUs. There is no intrinsic feature of the model other than the number of counts for each unit-level that suggest at what level the data should be combined to provide the most relevant archaeological understanding. The model provides largely similar results when aggregated at the EU or RU level, but with signficantly more uncertainly at the lower level of aggegation. However, apparent contradictions in the RCD dating at the EU-level -- where lesser depths yield earlier dates -- are sometimes resolved by disaggregating to the RU-level (Figure S8). Aggregation at the site level leads to fewer inferred CPs and markedly less discrimination among the sites, showing the pitfalls of overaggregation (Figure S6). However, these observations may be specific to the context of the Mandara mountains and general guidance as to how to navigate this tension will only be possible through corroboration with other artifact types and the methods application to a range of archaeological contexts.

\subsection*{Acknowledgements}
The authors are grateful to Gabrielle Grabin, Ana Lagunez, and Ryan Knauss for careful reviews and thoughtful discussion of the manuscript.

\bibliography{cp}

\begin{figure*}[!ht]
\begin{center}
\includegraphics[scale=0.32]{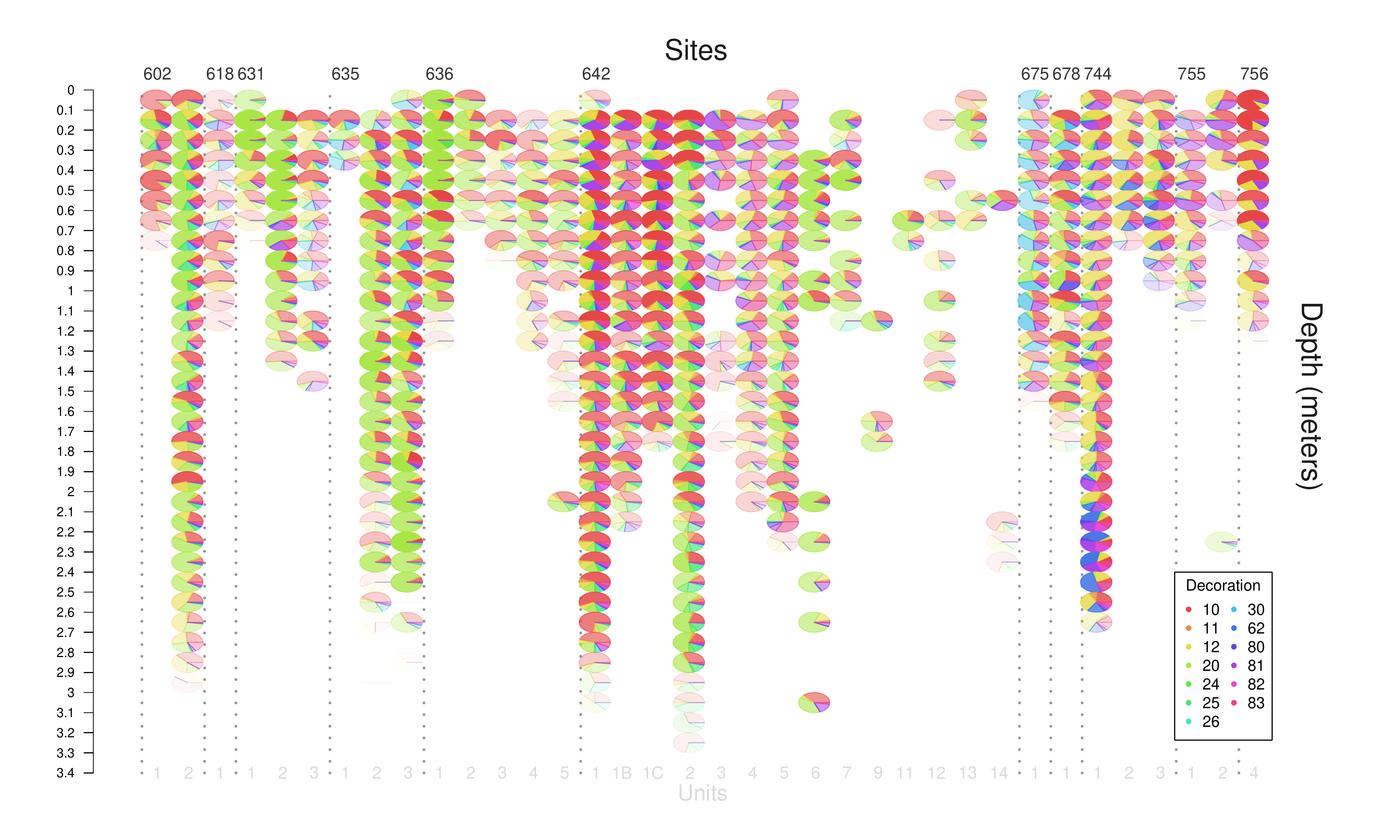}
\end{center}
\caption{Summary of observed distribution of decoration types present within each unit-level. Shading is logarithmic function of the number of observations within a unit-level, with lighter shades indicating fewer observations. }
\label{raw}
\end{figure*}

\begin{figure*}
\begin{center}
\includegraphics[scale=0.32]{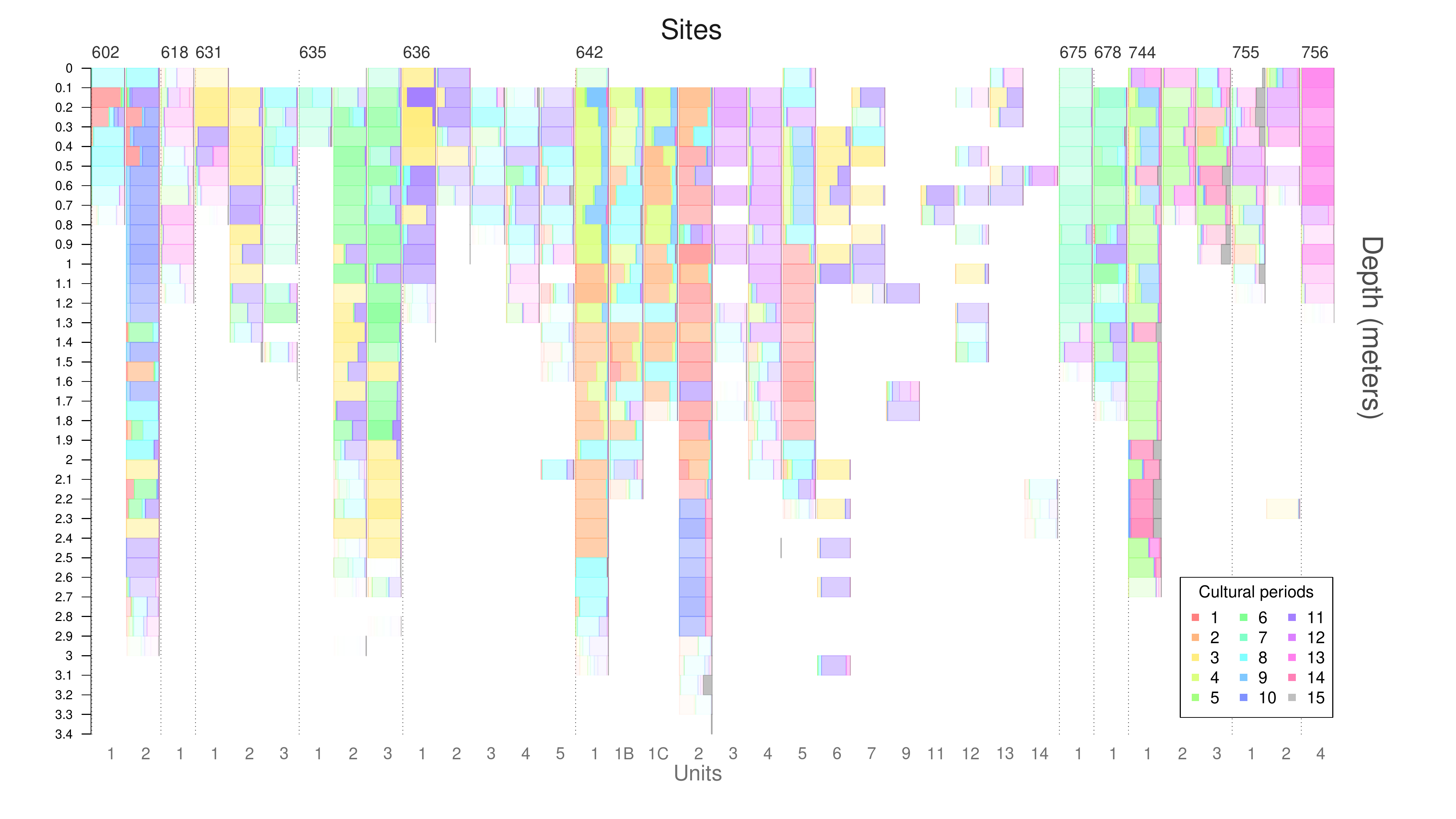}
\end{center}
\caption{Culture painting summary of CP posterior distribution for each unit-level. Coloring with 14 CPs present, with CP 15 aggregating additional epochs. Shading is the same as in Figure \ref{raw}.}
\label{painting}
\end{figure*}

\begin{figure}[!ht]
\begin{center}
\includegraphics[scale=0.3]{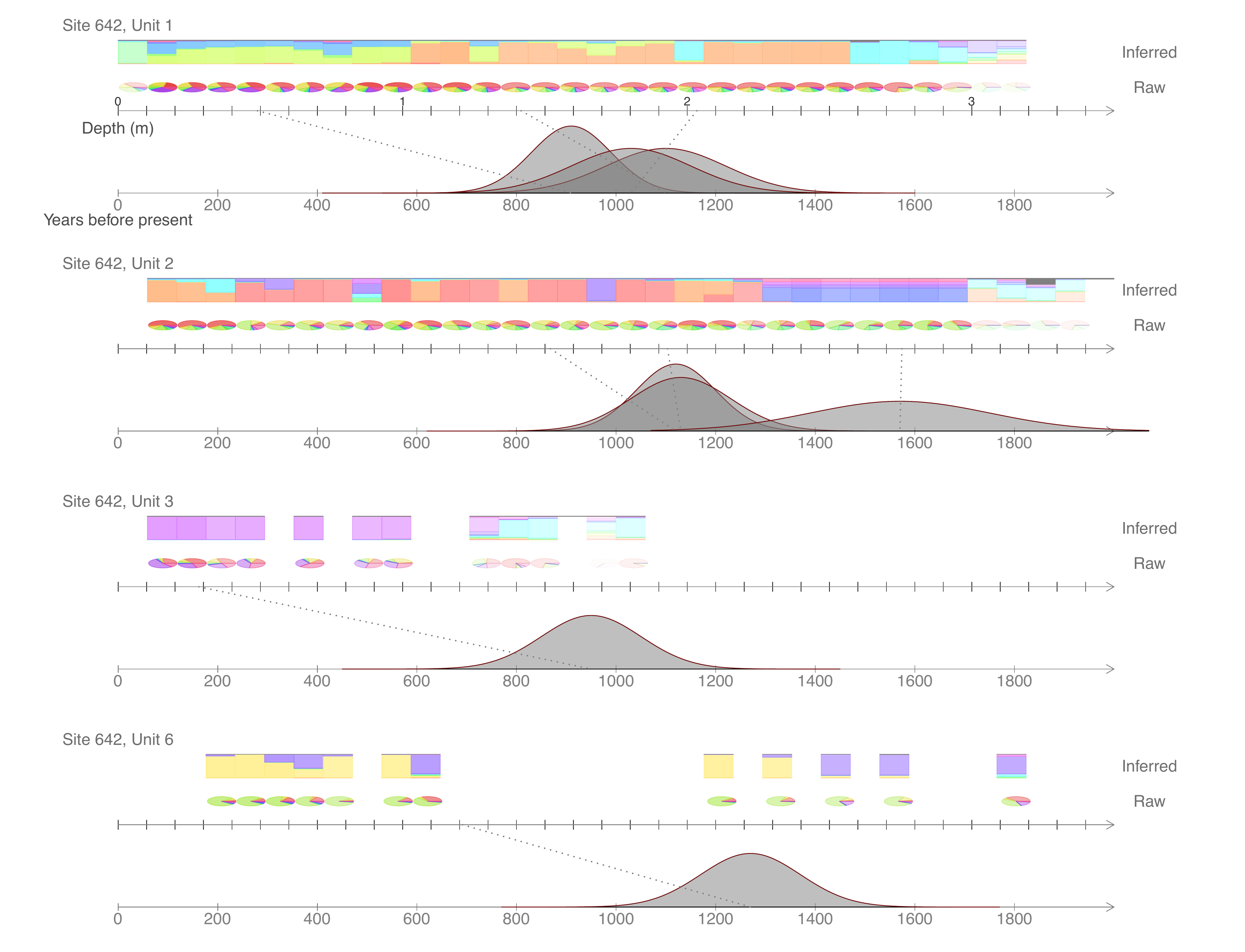}
\end{center}
\caption{Comparing RCD information with culture painting for site 642. For each of the four tested EUs, the panel contains: the RCD-estimated time before present (bottom) with error distribution; the depth location for each artifact (middle dashed line, connected by dotted line); the raw decoration type distributions; and the inferred culture painting. Color schemes are identifcal to Figures \ref{raw} and \ref{painting}, respectively. }
\label{rcd}
\end{figure}

\begin{center}
\textbf{SUPPLEMENTARY INFORMATION}
\end{center}

\section*{Methods}

\subsection*{Model}

The distribution $\mbox{\small{DP}}(G_0,\gamma) $ hides a great deal of complexity, and the model may be more intuitively considered as the infinite limit $K\rightarrow\infty$ of a finite mixture model given by
\begin{eqnarray*}
s_{ij} | c_{ij}=k, \theta &\sim & \mbox{\small{DM}}(A_k) \\
c_{ij} | p_1,\cdots,p_K  & \sim &\mbox{\small{DISCRETE}}(p_1, \cdots,p_K) \\
A_k & \sim & G_0 \\
p_1,\cdots,p_K &\sim & \mbox{\small{DIRICHLET}}(\gamma/K,\cdots,\gamma/K)
\end{eqnarray*}
The base measure and $\gamma$ are the same as before. This presentation makes clear the dependency on $c_{ij}$, the CP assignment of unit-level $(ij)$ that we retain as part of the inference procedure. 

\subsection*{Inference}
We employ a standard Markov chain Monte Carlo (MCMC) approach to sampling the posterior distribution of the model, closely following the methods described in \cite{Neal2000, Stein2013}. We employ a set of Gibbs updates, using a collapsed Gibbs sampling update step for the latent categorical variables in the DPM and an efficient Gibbs sampling routine for the DM parameters. At each iteration, the collapsed Gibbs update successively moves the unit-level latent variables to (possibly) new states, in the following way that closely follows Algorithm 8 from \cite{Neal2000}. For each unit-level $(i,j)$, let $l$ be the number of distinct culture labels $c_{(r,s)}$ for $(r,s) \neq (i,j)$. Let $h = l + m$ where $m$ is a parameter that allows for the number of new cultures for possible assignment. Here, we set $m=3$. If $c_{ij} = c_{rs}$ for some $(r,s)=(i,j)$, we draw values independently from the base measure for the corresponding component. If not (that is, $(i,j) \neq (r,s)$ for any $(r,s)$), then we set $c_{ij} = l+1$ and draw component values for each integer value between $l+1$ and $l+m$. Then we sample a new value for $c_{i,j}$ according to

where $c_{-(ij)}$ denotes all values of $c$ except $c_{ij}$, $n_{-(ij),c}$ is the number of values equal to $c$ for $(i,j) \neq (r,s)$, and $N$ is the total number of unit-levels. At each iteration, the set of $A_c$'s is purged of components that are not associated with any unit-level and appropriately renumbered. The Gibbs steps to update the DM parameters derive from a recently introduced approach for the efficient sampling of DM models. This employs a data augmentation scheme that creates an auxillary set of variables that correspond to an unobserved double replacement scheme on the multinomial counts. This augmentation, laid out together with the entire Gibbs sampling scheme in \cite{Stein2013}, permits the separation of the DM model into a Dirichlet distribution that is conjugate to the likelihood and so sampled directly, and a single parameter sampled via a griddy Gibbs update \cite{Ritter1992}.

\subsection*{Implementation and post-processing}

The algorithm is implemented in a set of scripts in the R computing environment \cite{Team2000}. These are freely available under a Creative Commons license and can be downloaded from the Digital Archaeological Archive (Document 400759) as well as the cleaned data set (Document 400758). Scripts for filtering the raw data set into the cleaned data set described in the Data subsection and generating the visualizations used in the manuscript are available upon request. 

We ran ten iterations of the model for one million iterations with the unit-level EU data with $30\%$ burn-in, taking about two weeks on a high-performance computing cluster running the Unix operating system CentOS 5. Visual examination the output indicated strongly similar results for each run. To ascertain convergence to the stationary distribution, we performed a number of posterior checks on the Markov chains, including Geweke's convergence diagnostic, examination of likelihood plots across the chain, and calculation of effective sample size for the likelihood. Thinning by 100 iterations reduces autocorrelation to acceptable levels, yielding an effective sample size of 265.31 for the run presented in the manuscript. Geweke's diagnostic for the same run yields a Z-score of 1.412, indicating a reasonably similar distribution between sections of the chain.

To visually summarize the data across the posterior sample, we need to bring the CP assignments at each iteration to be in correspondence with each other. Since the CPs are not intrinsically identifiable, this leads to a label switching problem, which has no general resolution for DPMs. We adapt the Kullback-Leibler (KL) divergence minimization method for addressing this problem in the context of finite mixture models \cite{Stephens2000} to the current situation, leveraging specific observations about the observed posterior distribution to make this approximation. We first consider the incidence plot across the chain, shown in Figure \ref{pairwise}, showing the frequency that each unit-level occupies the same CP as each other unit-level, shaded blue-to-red with increasing indicidence. We further observe that, after applying a $k$-mediods algorithm with fourteen CPs on the incidence matrix (Figure \ref{pairwise}, right side), there are fourteen well-defined CPs with the fifteenth CP distinguished by site-levels that associate with a variety of other clusters. Increasing and decreasing the number of clusters did not appreciably change this effect. We also observe that those iterations with fifteen or more clusters contain clusters that are significantly smaller than the remaining clusters ($<$10 members). For the purpose of visualization, this allows us to effectively reduce the model to the finite case by altering the chain labeling so that the smallest clusters for iterations with more than fourteen CPs are relabeled to be included in a single cluster fifteen. We then apply the KL divergence method to the fourteen primary CPs to provide consistent labeling across the chain. This produces a very similar result to a painting based on a finite-mixture model with 14 CPs.

\section*{Results}

\subsection*{Data from Mandara mountain sites}

We plot the frequency of different numbers of CPs inferred from the posterior MCMC samples in Figure \ref{pairwise} showing a mininum of 14 CPs. The modal number of CPs is 16, although the additional two CPs are nearly always highly variable unit-levels and comparatively small counts ($<$ 10 members).

\subsubsection*{Inferred distribution of potsherd production by culture}

\begin{figure*}[!ht]
\begin{center}
\includegraphics[scale=0.3]{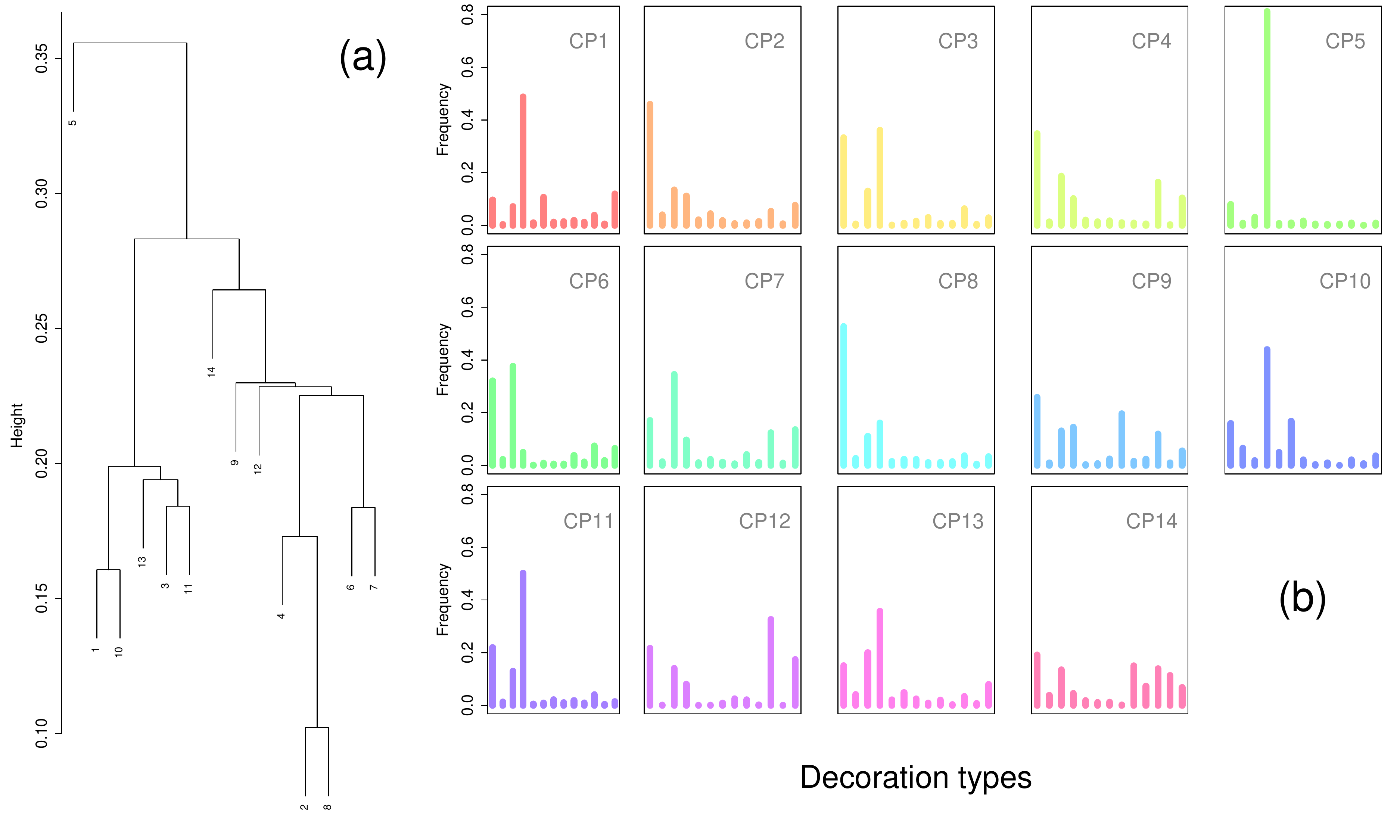}
\end{center}
\caption{Visualizations of DM model parameters. (a) Hierarchical clustering of the Euclidean distance among DM parameters for the fourteen largest CPs in the maximum likelihood posterior sample. (b) Expected frequency of the thirteen decoration types for each CP for the same iteration, with colors the same as in Figure \ref{painting}. }
\label{decoration}
\end{figure*}

To understand the relationship between the CPs, we visualize the inferred DM parameters from the maximum likelihood iteration in Figure \ref{decoration}. Figure (a) shows a hierarchical clustering of the parameters for each CP, with the distance between each calculated using a Euclidean metric and the single-linkage method for clustering.  The CPs associate in ways that suggest a process of cultural progression, with CPs often found at lower depths pairing closely with other CPs found a more shallow levels.

\subsection*{Culture painting by excavation unit}
As an aid to the reader, we present the culture painting from the MCMC run presented in Figure \ref{site_painting} but with additional shading indicating the relative number of observations.

\subsubsection*{Culture painting for site-aggregated data}

\begin{figure*}[!ht]
\begin{center}
\includegraphics[scale=0.6]{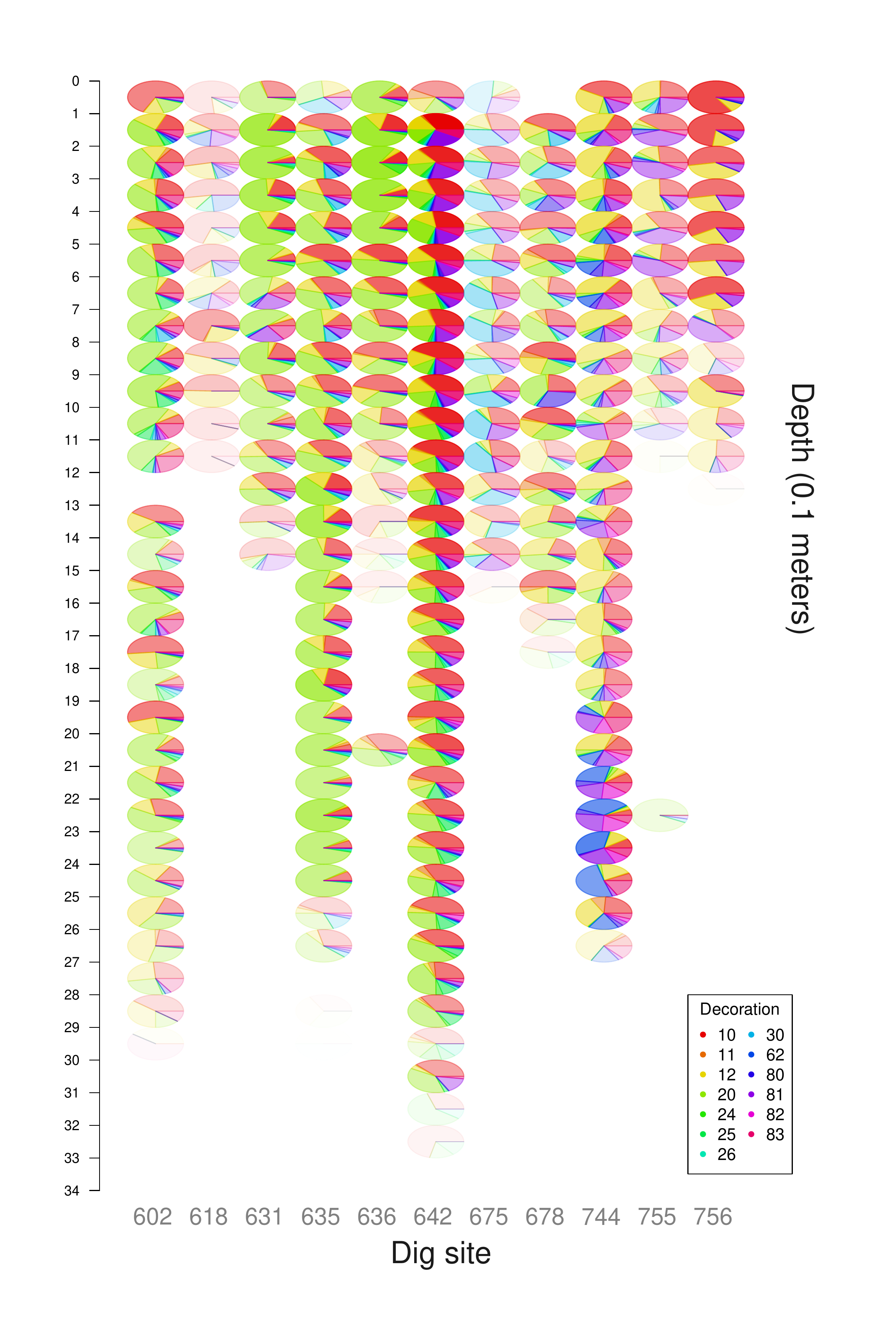}
\end{center}
\caption{Raw distribution of external decoration types aggregated by site. Colors denote decoration type; shading denotes the relative frequency of observations within a level.}
\label{site_raw}
\end{figure*}

\begin{figure*}[!ht]
\begin{center}
\includegraphics[scale=0.7]{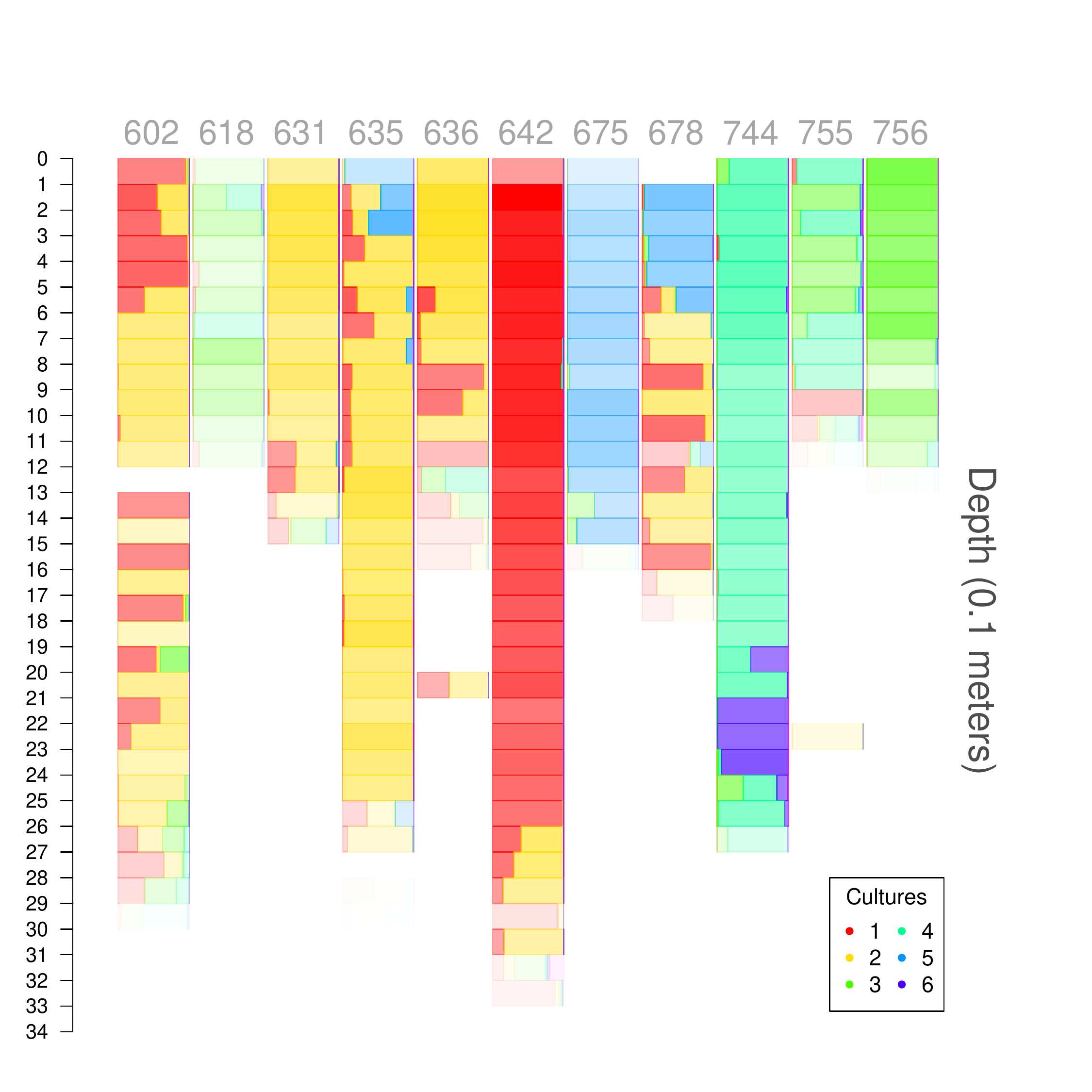}
\end{center}
\caption{Culture painting from site-aggegrated data set shows six cultures, with strong correlations across depths and distinct distributions by site.}
\label{site_painting}
\end{figure*}

Aggregating the data set at the site level yields 221 non-zero unit-levels, as shown in Figure \ref{site_raw}. Applying the culture painting algorithm to this data set yields the results presented in Figure \ref{site_painting}. The algorithm indicates a minimum of 6 CPs, with strong correlations across depths and clear distinctions across sites. As in the EU-level aggregation, we observe that the Neolithic sites exhibit the same CP. This CP is nearly absent from other sites, except for Site 755, that exhibit partial membership in this CP. Excluding the Neolithic sites, the painting indicates a less strict segregation between the eastern and main sites, with Sites 602 showing a mixture of CPs 1 and 2, also found in Sites 631, 635, 678, and 641. The large number of switchbacks and indistinct membership at this site possily indicate that the model is attempting to cope with insufficient data.

\subsubsection*{Culture painting by recording unit}

\begin{figure*}[!ht]
\begin{center}
\includegraphics[scale=0.2]{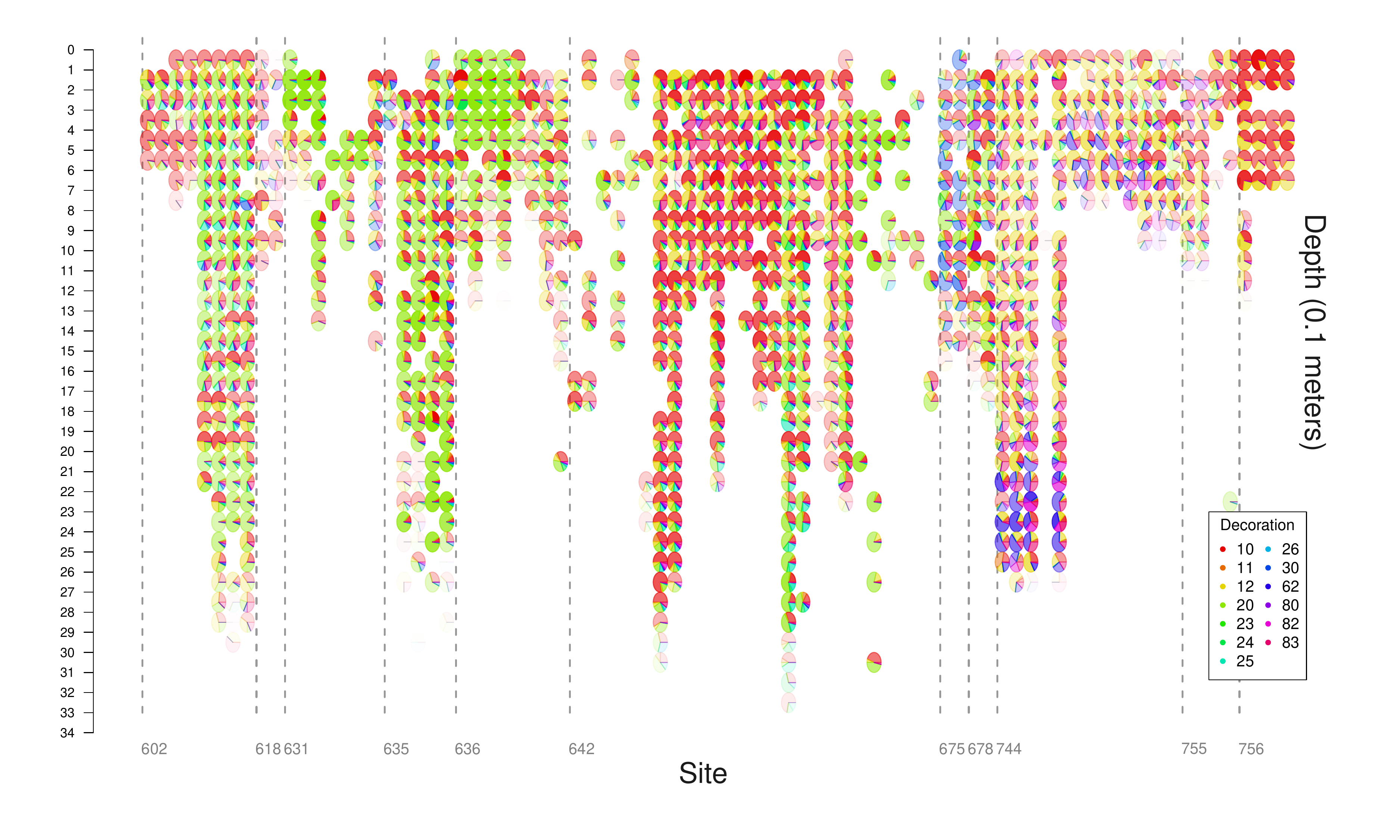}
\end{center}
\caption{Raw distribution of external decoration types aggregated by recording unit. Colors denote decoration type; shading denotes the relative frequency of observations within a level. }
\label{dig_raw}
\end{figure*}

\begin{figure*}[!ht]
\begin{center}
\includegraphics[scale=0.15]{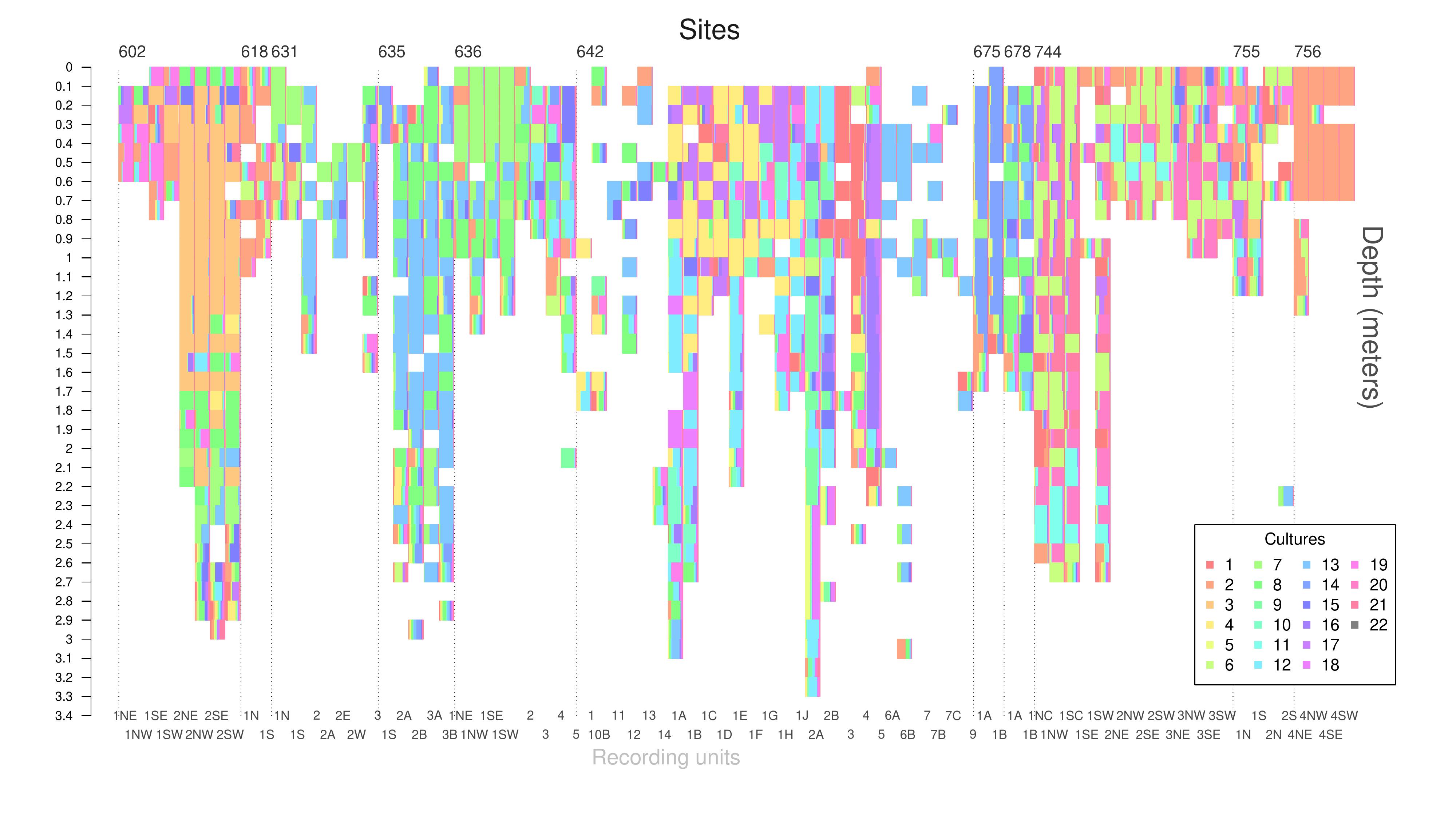}
\end{center}
\caption{Culture painting from recording-unit-aggegrated data set shows 21 cultures. To highlight differences between CPs, the shading used in previous painting plots is not used. }
\label{dig_painting}
\end{figure*}

Aggregating the data set at the RU level yields 1021 non-zero unit-levels, as shown in Figure \ref{dig_raw}. Applying the culture painting algorithm to this data set yields the results presented in Figure \ref{dig_painting}. The model indicates a minumum of 21 CPs. We continue to observe that the Neolithic sites are dominated by a single CP (here, CP 2). We also observe small amounts of CP 2 at the surface of site 755 and deep within sites 636, 642, and 744. Excluding the Neolithic sites, we observe strong segregation by CP between the eastern and main sites, with the former dominated by CPs 3, 6, 8, 17 and 21, with smaller introgressions from CPs 2, 3, and 7. The main sites are dominated by CPs 1, 2, 4, 7, 10, 11, 12, 13, 14, 15, 16, 17, 19, and 20. 

Despite the disaggregation to the RU level, we continue to observe strong segragation between the unit-levels within Site 642, with recording units (RUs) within excavation unit (EU) 1 exhibiting CPs 4, 12, and 16, RUs within EU 2 and 3 exhibiting largely CP 1, RUs within EU 4 showing CPs 15 and 16, and RUs within EUs 6-11 indicating membership largely in CPs 7 and 14. This division, even with substantially less data per unit-level,  further supports the position of historical ethnolinguistic segregation. The disaggregation indicates long-term segregation within sites 602 and 636. 

\subsection*{Simulations}
\begin{figure*}[!ht]
\begin{center}
\includegraphics[scale=0.4]{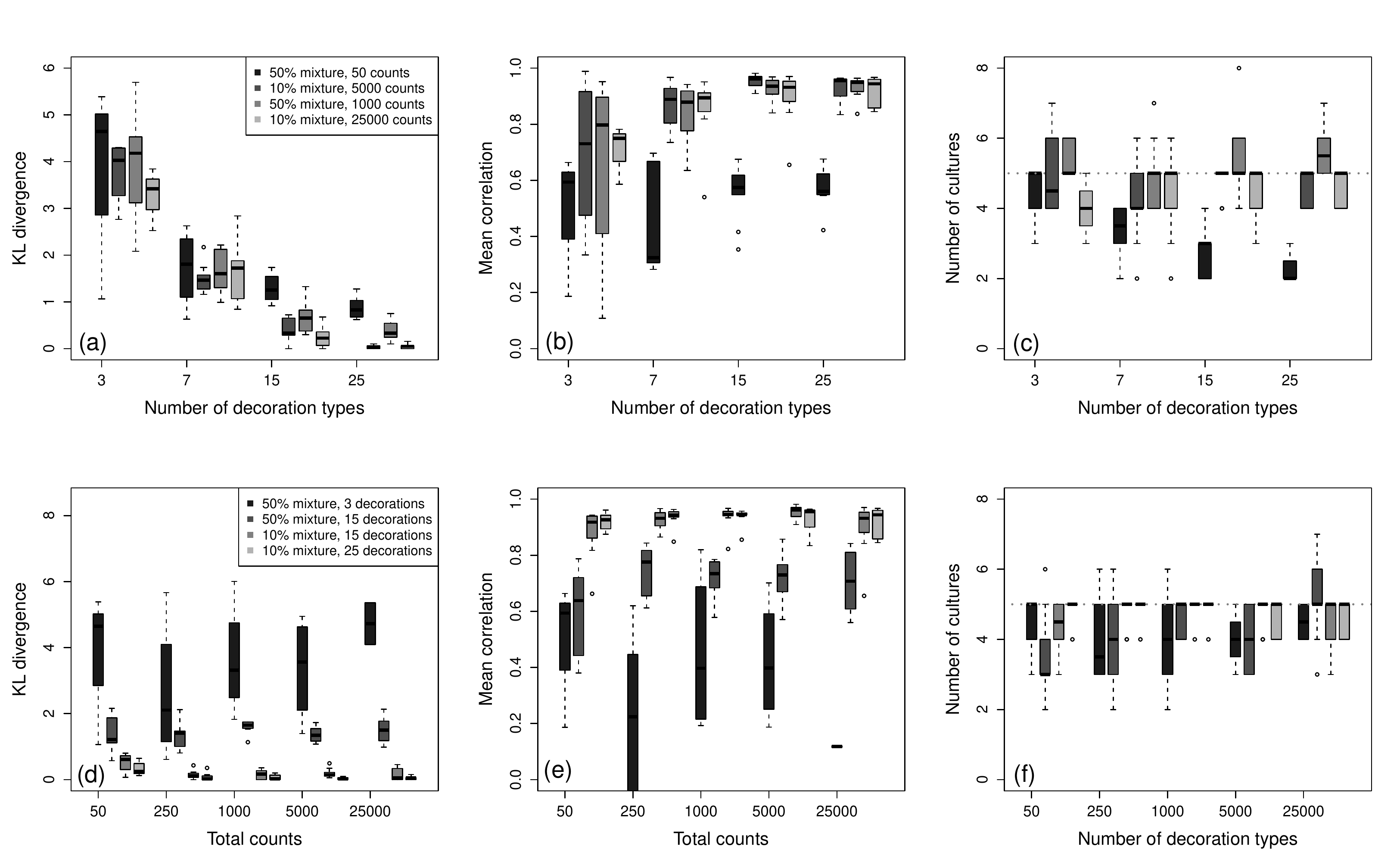}
\end{center}
\caption{Summary of simulation results for representative parameter regimes. }
\label{simulations}
\end{figure*}

Supplementary Table \label{parameters} shows the parameter values used in the simulation study, with each parameter set run with 10 iterations of the algorithm for 100,000 iterations. Figure \ref{simulations} summarizes the results for a set of representative parameter choices. The left panels show the KL divergence between the inferred and simulated CP assignments summed across all unit-levels. The middle panel shows the mean correlation between the maximum likelihood iteration and the simulated value, averaged over the best fit pairing for each of the simulated parameter vectors. The right panels show the modal number of inferred clusters, discarding clusters with fewer than five members. The study indicates that the number of decoration types is the strongest determinant of model performance, with the number of counts also playing a strong role. The model is less sensitive to mixture at moderate levels, but can be cofounded when the level of autocorrelation is low and degree of mixture is high ($50\%$ of counts).

\begin{table}[!ht]
\begin{center}
$\mbox{\small{PARAMETER VALUES FOR SIMULATION STUDY}}$ \\
\vspace{0.5cm}
\begin{tabular}{l|l}
Parameter & Values \\ 
\hline \hline 
Number of decoration types & 3, \ 7, \  15, \ 25 \\
Number of counts & 50, \ 250, \ 1,000, \ 5,000, \ 25,000 \\
Autocorrelation between levels & 0, \ 0.1, \ 0.5 \\
Mixture proportion & 0.0, \ 0.1, \ 0.5 \\
\end{tabular}
\caption{Table of parameters for simulation study. Each parameter set was run for ten iterations.}
\label{parameters}
\end{center}
\end{table}

\beginsupplement

\begin{figure*}[!ht]
\begin{center}
\includegraphics[scale=0.4]{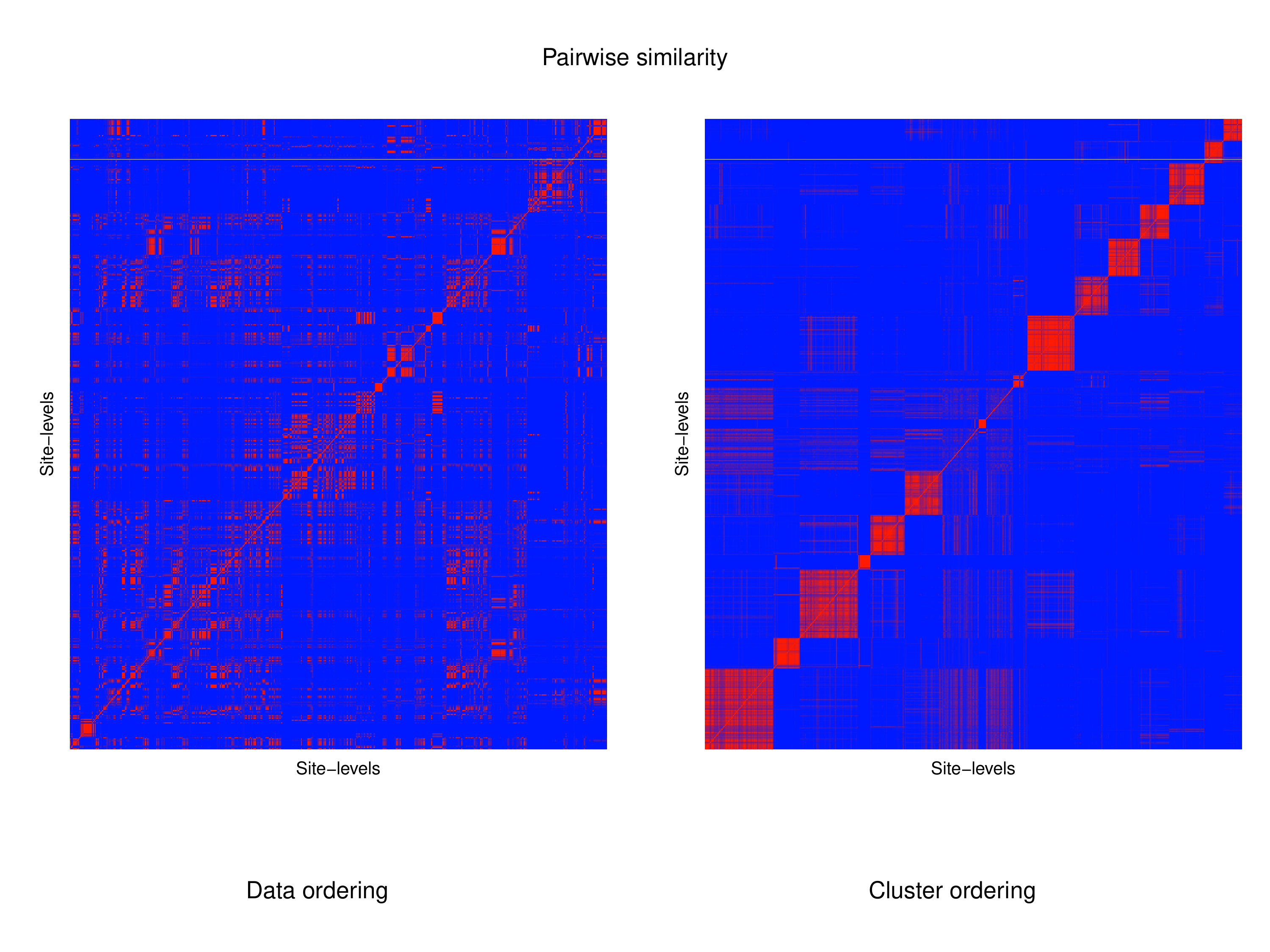}
\end{center}
\caption{Pairwise incidence for each of the 514 unit-levels with non-zero counts, with original data ordering on left and clustered ordering on right.}
\label{pairwise}
\end{figure*}

\begin{figure*}[!ht]
\begin{center}
\includegraphics[scale=0.3]{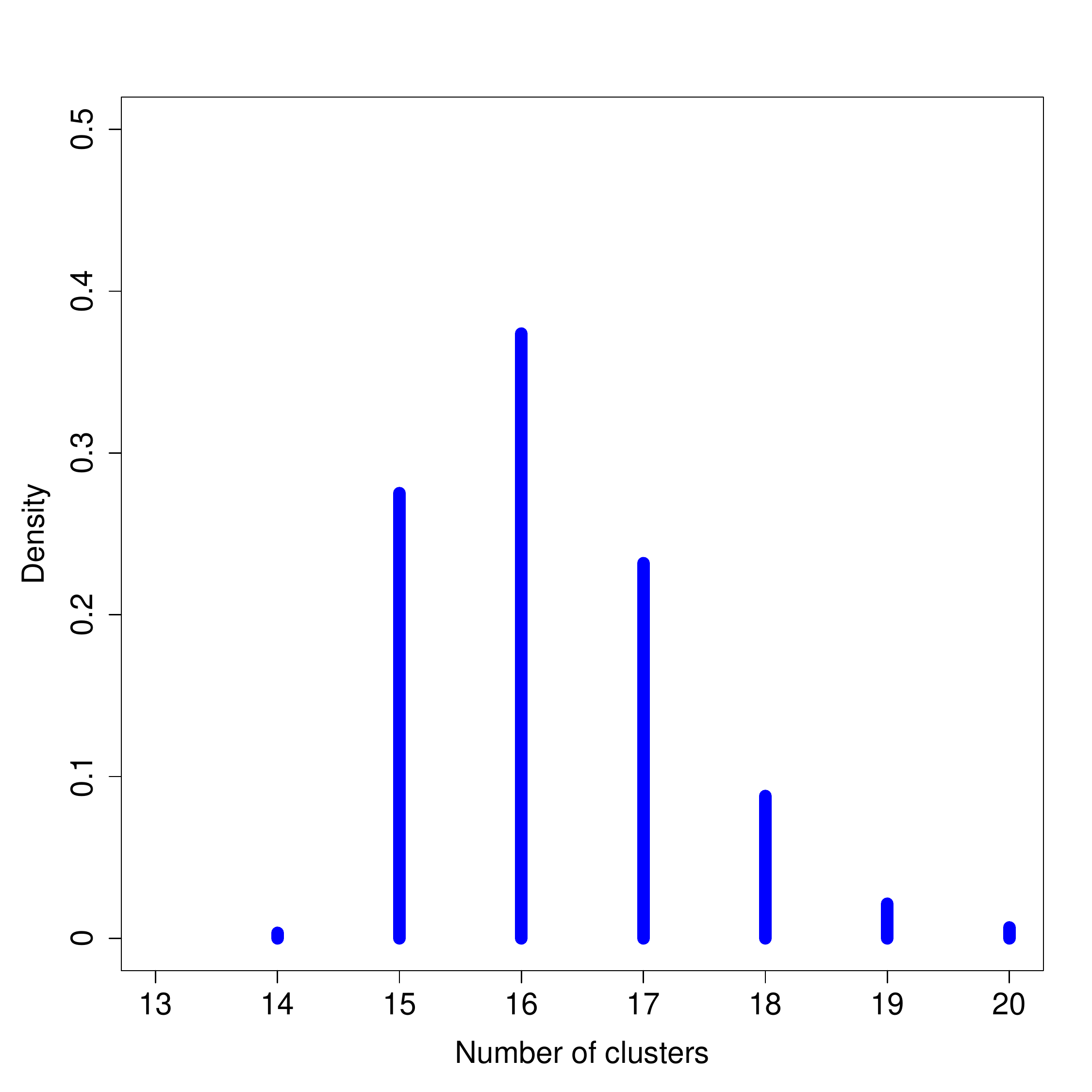}
\end{center}
\caption{Frequency of different number of CPs in the posterior distribution for the run presented in the manuscript, excluding 30\% burn-in.}
\label{pairwise}
\end{figure*}

\begin{figure*}[!ht]
\begin{center}
\includegraphics[scale=0.3]{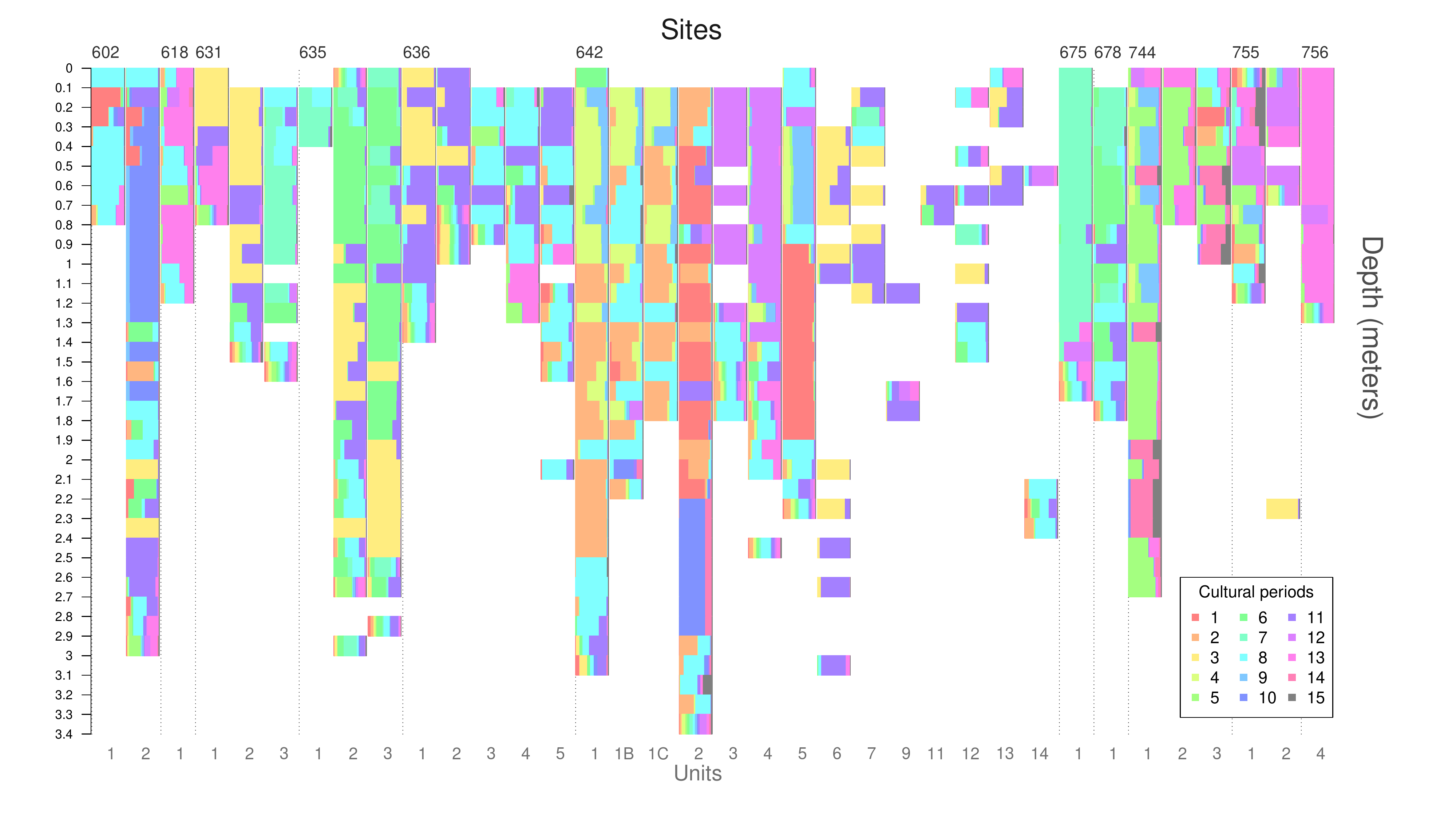}
\end{center}
\caption{Posterior summary of culture painting from unit-aggegrated data set without shading. The plot is otherwise identical to Figure 3. }
\label{shaded}
\end{figure*}


\end{document}